\documentclass[11pt, a4paper]{article}
\usepackage{amsmath}

\newtheorem{theorem}{Theorem}

\newcommand{\benumerate}{\begin{enumerate}}
\newcommand{\eenumerate}{\end{enumerate}}

\newcommand{\bitemize}{\begin{itemize}}

\newcommand{\eitemize}{\end{itemize}}
\newcommand{\ep}{\epsilon}

\begin{document}

\title{Integrable equations in $2+1$ dimensions: deformations of dispersionless limits}
\author{E.V. Ferapontov, A. Moro and V.S. Novikov}
    \date{}
    \maketitle
    \vspace{-7mm}
\begin{center}
Department of Mathematical Sciences \\ Loughborough University \\
Loughborough, Leicestershire LE11 3TU \\ United Kingdom \\[2ex]
e-mails: \\[1ex] \texttt{E.V.Ferapontov@lboro.ac.uk}\\
\texttt{A.Moro@lboro.ac.uk}\\
\texttt{V.Novikov@lboro.ac.uk}
\end{center}

\bigskip

\begin{abstract}

We classify  integrable  third order  equations in $2+1$
dimensions which generalize the examples of
Kadomtsev-Petviashvili, Veselov-Novikov and Harry Dym equations.
Our approach is based on the  observation that dispersionless
limits of integrable systems in $2+1$ dimensions possess
infinitely many  multi-phase solutions coming from the so-called
hydrodynamic reductions.
In this paper we adopt a novel perturbative approach to the
classification problem. Based on the method of hydrodynamic
reductions, we first classify integrable quasilinear systems which
may (potentially) occur as dispersionless limits of   soliton
equations in $2+1$ dimensions. To reconstruct  dispersive
deformations, we require that all hydrodynamic reductions of the
dispersionless limit are inherited by the corresponding dispersive
counterpart. This procedure leads to a complete list of integrable
third order equations, some of which are apparently new.

\bigskip


\bigskip

\end{abstract}

\newpage

\section{Introduction}

The classification of integrable  systems has been a topic of
active research from the very beginning of  soliton theory. In
$1+1$ dimensions, this resulted in extensive lists of integrable
equations within particularly important subclasses \cite{Mik0},
which were obtained by means of the symmetry approach. Although
this technique generalizes to $2+1$ dimensions, one encounters
additional difficulties due to the appearance of non-local
variables \cite{Mik1}. A way to bypass the problem of
non-locality, known as the  perturbative symmetry approach
\cite{Mik2},  provides an efficient way to classify soliton
equations in $2+1$ dimensions. In this framework, one starts with
a linear equation having degenerate dispersion law \cite{Shulman},
and reconstructs the allowed nonlinearity. However, few
classification results have been obtained so far. In fact, most of
the $(2+1)$-dimensional examples known to date were derived  by
postulating a  special structure of the corresponding Lax pair,
see e.g. \cite{Kon4},  \cite{Wang}.

In this paper we  adopt a novel approach to the problem of
classification of scalar third order soliton equations in $2+1$
dimensions with the `simplest'  possible non-localities,
$$
u_t=F(u, w, Du, Dw),
$$
and
$$
u_t=F(u,  v, w, Du, Dv, Dw),
$$
respectively. Here $u(x, y, t)$ is a scalar field, and  the
non-local variables $v(x, y, t)$ and $w(x, y, t)$ are defined via
$w_x=u_y$ and $v_y=u_x$, equivalently, $w=D_x^{-1}D_yu$, \
$v=D_y^{-1}D_xu$.  The symbols  $Du,  Dv,  Dw$ denote the
collection of all partial derivatives of $u, v, w$ with respect to
$x$ and $y$ up to the third order. In fact, it is sufficient to
allow only $y$-derivatives of $w$ and $x$-derivatives of $v$. We
will refer to the above equations as  the `non-symmetric' and
`symmetric' cases, respectively.  We  assume that in both cases
the dependence of the right hand side $F$ on the derivatives of
$u$ and $ w$ (resp, $u, v, w$) is {\it polynomial}, where the
coefficients are allowed to be arbitrary functions of  $u$ and $w$
(resp, $u, v, w$). Explicitly, in the non-symmetric case  we have
\begin{equation}
\label{nonsym}
\begin{aligned}
u_t&=\varphi u_x+\psi u_y +\eta w_y+\epsilon(...)+\epsilon^2(...),
~~~~ w_x=u_y,
\end{aligned}
\end{equation}
where  $\varphi, \psi, \eta$ are functions of $u$ and $w$, while
the terms at $\epsilon $ and $\epsilon^2$ are assumed to be
homogeneous differential polynomials of  the order two and three
in the derivatives of $u$ and $w$, whose coefficients  can be
arbitrary functions of $u$ and $w$. We use the following weighting
scheme: $u$ and $w$ are assumed to have order zero, their
derivatives $u_x, u_y, w_x, w_y$ are of order one, the expressions
$u_{xx}, u_{xy}, u_{yy}, w_{yy}, u_x^2, u_xu_y, u_y^2, u_xw_y,
u_yw_y, w_y^2$ are of order two, etc. Thus, the term at $\epsilon$
is a linear combination of the ten second order expressions whose
coefficients  can be arbitrary functions of $u$ and $w$. The most
familiar example within the class (\ref{nonsym}) is the
Kadomtsev-Petviashvili (KP) equation,
$$
\begin{aligned}
u_{t}& = u u_{x} +w_y+ \epsilon^{2} u_{xxx}, ~~~~ w_x=u_y.
\end{aligned}
$$
Similarly, in the symmetric case  we consider equations of the
form
\begin{equation}
\label{sym}
\begin{aligned}
u_t&=\varphi u_x+\psi u_y +\eta w_y+\tau
v_x+\epsilon(...)+\epsilon^2(...), ~~~~ w_x=u_y,~ v_y=u_x,
\end{aligned}
\end{equation}
here  $\varphi, \psi, \eta, \tau$   are functions of $u, v$ and
$w$. A  canonical example of the form (\ref{sym}) is the
Veselov-Novikov (VN) equation,
$$
\begin{aligned}
u_{t} &=(uv)_x+(u w)_{y} +\epsilon^2 ( u_{xxx}+ u_{yyy}), ~~~~
w_x=u_y,~ v_y=u_x.
\end{aligned}
$$
In Sect.\ 2 we bring together other known examples of the form
(\ref{nonsym}) and (\ref{sym}) which include the KP,  VN, Harry
Dym  equations and their modifications.

Our approach to the classification problem is based on the
following key observations.

\begin{itemize}
\item {\bf Dispersionless limits of integrable soliton equations
in $2+1$ dimensions possess infinitely many hydrodynamic
reductions.  }
\end{itemize}

\noindent  In particular, dispersionless limits of Eqs.\
(\ref{nonsym}) and (\ref{sym}),
\begin{equation}
\label{dnonsym}
\begin{aligned}
u_t&=\varphi u_x+\psi u_y +\eta w_y, ~~~~~ w_x=u_y,
\end{aligned}
\end{equation}
and
\begin{equation}
\label{dsym}
\begin{aligned}
u_t&=\varphi u_x+\psi u_y +\eta w_y+\tau v_x, ~~~~ w_x=u_y, ~
v_y=u_x,
\end{aligned}
\end{equation}
should  possess infinitely many hydrodynamic reductions and, thus,
must be integrable in the sense of  \cite{Fer4}. It was observed
in  \cite{Fer4} that the method of hydrodynamic   reductions
provides an efficient classification criterion. Thus, as a first
step, in Sect.\ 3 we classify integrable  first order equations of
the form (\ref{dnonsym}) and (\ref{dsym}) which may (potentially)
occur as dispersionless limits of integrable equations of the form
(\ref{nonsym}) and (\ref{sym}). We emphasize that the requirement
of being a
 dispersionless limit of a {\it third order} soliton equation imposes further severe constraints, so that very few particular cases obtained in Sect.\ 3 do actually survive.

 Given an integrable dispersionless equation, one needs to reconstruct  dispersive deformations. In $1+1$ dimensions, this problem has been a subject of extensive research in \cite{Dub1, Dub2, Dub3, Zhang}, see also \cite{Baikov}. In $2+1$ dimensions, the reconstruction procedure is  based on the following key observation \cite{FerM}:

\begin{itemize}
\item {\bf Hydrodynamic reductions of dispersionless limits of
integrable soliton equations can be deformed into reductions of
the corresponding dispersive counterparts (strictly speaking, this
is only true if the  dispersionless limit is linearly
non-degenerate, see Sect. 4). Furthermore, the requirement of the
inheritance of all hydrodynamic reductions allows one to
efficiently reconstruct dispersive terms in $2+1$ dimensions. }
\end{itemize}

\noindent This suggests the following alternative {\bf definition}
of the integrability:

\medskip

\noindent {\it A $(2+1)$-dimensional system is said to be
integrable if all hydrodynamic reductions of its dispersionless
limit (which is assumed to be linearly non-degenerate) can be
deformed into reductions of the corresponding dispersive
counterpart.}

\medskip

\noindent Although this property is satisfied for all known
integrable equations whose dispersionless limit is not totally
linearly degenerate, it would be important to formulate more
precise statements about the equivalence of our definition with
more `conventional' approaches to the integrability.

The procedure of the reconstruction of dispersive terms  is
thoroughly illustrated  in Sect.\ 4, where we examine case-by-case
all integrable dispersionless limits from Sect.\ 3. Our
calculations result in a complete list of integrable
$(2+1)$-dimensional equations, some of which are apparently new.
It is important to emphasize that, although our approach is based
on the requirement of the inheritance of hydrodynamic reductions,
all examples from the final list do actually possess conventional
Lax pairs. Altogether, we found  three new equations. One of them
is
\begin{equation}
u_t=(\beta w+\beta^2u^2)u_x-3\beta u u_y+w_y+\epsilon^2
[B^3(u)-\beta u_x B^2(u)], ~~~~ w_x=u_y, \label{new1}
\end{equation}
here $B=\beta uD_x- D_y$, $\beta$=const. It possesses the Lax pair
\begin{gather*}
\begin{aligned}
\psi_{xy} &= \beta u \psi_{xx}+\frac{1}{3\epsilon^2} \psi ,  \\
\psi_t &=\epsilon^2\beta^3 u^3\psi_{xxx} -
\epsilon^2\psi_{yyy}+3\epsilon^2\beta^2uu_y\psi_{xx}+\beta
w\psi_x.
\end{aligned}
\end{gather*}
The second example is
\begin{equation}
u_t=\frac{4}{3}\beta^2 u^3u_x+(w-3\beta u^2) u_y+uw_y+\ep^2
[B^3(u)-\beta u_xB^2(u)], ~~~~ w_x=u_y, \label{new2}
\end{equation}
here again $B=\beta u D_x-D_y$, $\beta$=const. The corresponding
Lax pair is
\begin{gather*}
\begin{aligned}
\psi_{xy} &=\beta u \psi_{xx}+\frac{1}{3\epsilon^2}u\psi ,  \\
\psi_t &=\epsilon^2\beta^3u^3\psi_{xxx}
-\epsilon^2\psi_{yyy}+3\epsilon^2\beta^2uu_y\psi_{xx}+\frac{\beta^2}{3}u^3\psi_x+w\psi_y+\beta
uu_y\psi.
\end{aligned}
\end{gather*}
We point out that similar  Lax operators  appeared in the context
of the $(2+1)$-dimensional Camassa-Holm equation \cite{Zenchuk}.
Our last example is a deformation of the Harry Dym (HD) equation,
\begin{equation}
 u_{t} =\frac{\delta}{u^3}u_x-2w  u_{y} +u w_y -\frac{\epsilon^2}{u}\left(\frac{1}{u}\right)_{xxx}, ~~~~ w_x=u_y,
\label{new3}
\end{equation}
for $\delta=0$ it reduces to the standard HD equation (Example 6
of Sect.\ 2.1). It has the Lax pair $L_t=[A, L]$  where
\begin{gather*}
\begin{aligned}
L &= \frac{\ep^2}{u^2} D_x^2+\frac{\ep}{\sqrt 3} D_y+\frac{\delta}{4u^2},  \\
A &=\frac{4\ep^2}{u^3 } D_x^3+\left(-\frac{6\ep^2
u_x}{u^4}+\frac{2\sqrt 3 \ep w}{u^2}\right)
D_x^2+\frac{\delta}{u^3} D_x+\left(-\frac{3\delta
u_x}{2u^4}+\frac{\sqrt 3 \delta w}{2\ep u^2}\right).
\end{aligned}
\end{gather*}
All three examples belong to the non-symmetric case. In the
symmetric case we have  no new equations apart from those listed
in Sect.\ 2.2. This leads to the following main result:

\begin{theorem} Equations (\ref{new1}) -- (\ref{new3}) along with the known
examples of KP, non-symmetric VN, HD equations and their
modifications provide a complete list of integrable equations of
the form (\ref{nonsym}) with $\eta \ne 0$ whose dispersionless
limit is linearly nondegenerate:
\begin{align*}
 &{ KP ~ equation}& \qquad &
 u_{t} = u u_{x} +w_y{{+ \epsilon^{2} u_{xxx}}}, &\\
&{ mKP ~ equation}& \qquad &
u_{t} =(w- u^2/2)  u_{x} +w_y {{+\epsilon^2  u_{xxx}}},&\\
&{Gardner ~ equation}& \qquad &
 u_{t} =(\beta w- \frac{\beta^2}{2}u^2+\delta u)  u_{x} +w_y {{+\epsilon^2  u_{xxx}}},&\\
&{  VN ~ equation}& \qquad &
u_{t} =(u w)_{y} {{+\epsilon^2  u_{yyy}}},&\\
&{ mVN ~ equation}& \qquad &
u_{t} =(u w)_y {{+\epsilon^2  \left( u_{yy}-\frac{3}{4}\frac{u_y^2}{u}\right)_y}},&\\
&{HD ~ equation }& \qquad &
u_{t} =-2w  u_{y} +u w_y {{-\frac{\epsilon^2}{u}\left(\frac{1}{u}\right)_{xxx}}},\\
&{ deformed ~ HD~ equation }& \qquad &
u_{t} =\frac{\delta}{u^3}u_x-2w  u_{y} +u w_y {{-\frac{\epsilon^2}{u}\left(\frac{1}{u}\right)_{xxx}}},&\\
&{ Equation ~ (\ref{new1}) }& \qquad & u_t=(\beta
w+\beta^2u^2)u_x-3\beta u u_y+w_y+\epsilon^2
[B^3(u)-\beta u_x B^2(u)],&\\
&{Equation ~ (\ref{new2}) }& \qquad & u_t=\frac{4}{3}\beta^2
u^3u_x+(w-3\beta u^2) u_y+uw_y+\ep^2 [B^3(u)-\beta u_xB^2(u)].&
\end{align*}
\noindent In the symmetric case there exist only two examples of
integrable equations of the form (\ref{sym}) with $\eta, \tau \ne
0$:
\begin{align*}
&{  VN ~ equation}& \qquad & u_{t} =(uv)_x+(u w)_{y} +\epsilon^2
u_{xxx}+\epsilon
^2 u_{yyy},&\\
&{ mVN ~ equation}& \qquad & u_{t} =(uv)_x+(u w)_y +\epsilon^2
\left( u_{xx}-\frac{3}{4}\frac{u_x^2}{u}\right)_x+ \epsilon^2
\left( u_{yy}-\frac{3}{4}\frac{u_y^2}{u}\right)_y.
\end{align*}
\end{theorem}

\noindent The proof is summarised in Sect.\ 4. Under the
substitution $w=0, \ u_y=0$ the equations (\ref{new1}),
(\ref{new2}) reduce to
$$
u_{t} = \epsilon^{2} \beta^{3} \left(u^{3} u_{xxx} + 3 u^{2} u_{x}
u_{xx} \right) + \beta^{2} u^{2} u_{x}
$$
and
$$
 u_{t} =
\epsilon^{2} \beta^{3} \left(u^{3} u_{xxx} + 3 u^{2} u_{x} u_{xx}
\right) + \frac{4}{3} \beta^{2} u^{3} u_{x},
$$
respectively. In this form, they have appeared in \cite{watanabe},
see also~\cite{MSS} and references therein. It was pointed out
(see e.g. \cite{Ibragimov2,MSS, Sakovich}) that there exist
differential substitutions bringing these equations to a constant
separant form. It would be interesting to find out whether  Eqs.\
(\ref{new1}) -- (\ref{new3}) are related to any of the known
soliton hierarchies: the main problem here is that the above
differential substitutions do not extend to $2+1$ dimensions in
any obvious way.

\noindent {\bf Remark 1.} The examples of VN and mVN equations
show that different $(2+1)$-dimensional equations may have one and
the same dispersionless limit.

\noindent {\bf Remark 2.} Our approach to the classification
problem does not apply to non-symmetric equations with $\eta=0$
(or symmetric equations with $\eta=\tau=0$). As we explain in
Sect.\ 3, these conditions are equivalent to the reducibility of
the dispersion relations of the corresponding systems
(\ref{dnonsym}), (\ref{dsym}). A familiar example within this
class is  the so-called `breaking soliton' equation,
$$
u_t=2wu_x+4uu_y-\epsilon^2u_{xxy}, ~~~~ w_x=u_y,
$$
see e.g. \cite{Bog}. Here  $\varphi=2w, \ \psi=4u, \ \eta=0$.
Equations of this type are not amenable to the method of
hydrodynamic reductions, and require an alternative approach.

\section{Known Examples}

\subsection{Non-symmetric case}

Here we bring together known  examples of soliton equations  whose
dispersionless limit is of the form  (\ref{dnonsym}).  The
relation $w_x=u_y$ will be automatically assumed whenever $w$
appears explicitly in the equation. Examples 1-6 list  third order
equations. Examples 7-10 correspond to    equations of order five,
or differential-difference equations.

\bigskip

\noindent{\bf Example 1.} The Kadomtsev-Petviashvili (KP)
equation,
\begin{equation}
\label{kp} u_{t} = u u_{x} +w_y+ \epsilon^{2} u_{xxx},
\end{equation}
arises in mathematical physics as a  two-dimensional
generalization of the KdV equation.  Its dispersionless limit (dKP
equation),
\begin{equation}
\label{dkp} u_{t} = u u_{x}+w_y,
\end{equation}
 also known as the Khokhlov-Zabolotskaya equation \cite{KZ}, is of interest in its own, playing important role in non-linear acoustics, gas dynamics and differential geometry.

 \bigskip

 \noindent{\bf Example 2.}
The modified KP  (mKP) equation,
\begin{equation}
\label{mkp} u_{t} =(w- u^2/2)  u_{x} +w_y +\epsilon^2  u_{xxx},
\end{equation}
has the dispersionless limit
\begin{equation}
\label{dmkp} u_{t} =(w- u^2/2)  u_{x} +w_y.
\end{equation}

\bigskip

 \noindent{\bf Example 3.}
The $(2+1)$-dimensional version of the Gardner equation is of the
form \cite{Kon4},
\begin{equation}
\label{mkp} u_{t} =(\beta w- \frac{\beta^2}{2}u^2+\delta u)  u_{x}
+w_y +\epsilon^2  u_{xxx},
\end{equation}
which reduces to the KP or mKP equations upon setting $\beta=0$ or
$\delta=0$, respectively. Its dispersionless limit has the form
\begin{equation}
\label{dmkp} u_{t} =(\beta w- \frac{\beta^2}{2}u^2+\delta u)
u_{x}+w_y.
\end{equation}

\bigskip

 \noindent{\bf Example 4.}
The non-symmetric version of the Veselov-Novikov  equation
\cite{VesNov, Nizhnik, Boiti},
\begin{equation}
\label{vn} u_{t} =(u w)_{y} +\epsilon^2  u_{yyy},
\end{equation}
has the dispersionless limit
\begin{equation}
\label{dvn} u_{t} =(u w)_y.
\end{equation}

\bigskip

 \noindent{\bf Example 5.}
The non-symmetric version of the modified Veselov-Novikov equation
\cite{Bogdanov},
\begin{equation}
\label{vn} u_{t} =(u w)_y +\epsilon^2  \left(
u_{yy}-\frac{3}{4}\frac{u_y^2}{u}\right)_y,
\end{equation}
has the same dispersionless limit as in the previous example,
\begin{equation}
\label{dvn} u_{t} =(u w)_y.
\end{equation}

\bigskip

  \noindent{\bf Example 6.}
The Harry Dym equation \cite{Kon4},
\begin{equation}
\label{vn} u_{t} =-2w  u_{y} +u w_y
-\frac{\epsilon^2}{u}\left(\frac{1}{u}\right)_{xxx},
\end{equation}
(set $\tilde u=1/u$ to obtain the equation from \cite{Kon4}), has
the dispersionless limit
\begin{equation}
\label{dvn} u_{t} =-2w  u_{y} +u w_y.
\end{equation}

\bigskip

\noindent{\bf Example 7.} The fifth order version of the Harry Dym
equation is
\begin{equation}
\label{5HD} u_{t} =-3w  u_{y} +u
w_y-\frac{\epsilon^2}{u^4}(u^2u_{xxy}-3u(u_xu_y)_x+6u_x^2u_y)
+\frac{\epsilon^4}{u^2}\left(\frac{1}{u^2}\right)_{xxxxx},
\end{equation}
see \cite{Kon4}. Its dispersionless limit has the form
\begin{equation}
\label{d5HD} u_{t} =-3w  u_{y} +u w_y.
\end{equation}

\bigskip

\noindent {\bf Example 8.} The Toda lattice  is a system of two
differential-difference equations
\begin{gather}
\label{TL}
\begin{aligned}
 \epsilon u_{t} &= u \; (w(y) - w(y-\epsilon)), \\
\epsilon w_{x} &= u(y+ \epsilon) - u(y),
\end{aligned}
\end{gather}
or
\begin{gather}
\label{TL_exp}
\begin{aligned}
u_{t}/u &= w_{y} - \frac{\epsilon}{2} w_{yy} +
\frac{\epsilon^{2}}{6} w_{yyy} + \dots + (-1)^{n+1}
\frac{\epsilon^{n}}{n!} w_{ny} + \dots, \\
w_{x} &= u_{y} + \frac{\ep}{2} u_{yy} + \frac{\ep^{2}}{6} u_{yyy}+
\dots + \frac{\ep^{n}}{n!} u_{ny} + \dots.
\end{aligned}
\end{gather}
Its dispersionless limit is
\begin{equation}
u_{t} = u w_{y}. \label{dTL}
\end{equation}

\bigskip

\noindent {\bf Example 9.} The nonlocal Toda lattice equation is
\begin{gather}
\label{nTL}
\begin{aligned}
\displaystyle  \ep \sigma_{xt} &= e^{ \frac{\sigma (x+\ep,
y+\ep)-\sigma}{\ep}}-e^{\frac{\sigma -\sigma (x-\ep,
y-\ep)}{\ep}},
  \end{aligned}
\end{gather}
see  \cite{Ustinov}. Its dispersionless limit is
\begin{equation}
\sigma_{xt} = e^{\sigma_x+\sigma_y}(\sigma _{xx}+2\sigma
_{xy}+\sigma_{yy}), \label{dTL}
\end{equation}
or, setting $\sigma_x=u, \ \sigma_y=w$,
$$
u_t=e^{u+w}(u_x+2u_y+w_y).
$$

\bigskip

\noindent{\bf Example 10.} The BKP and CKP equations are of the
form
\begin{gather}
\label{KK}
\begin{aligned}
u_{t} - 5 (u^{2} + w) u_{x} - 5 u w_{x} + 5 w_{y} + \epsilon^{2} (
u u_{xxx} +  w_{xxx} +  u_{xxx}) -\frac{ \epsilon^{4}}{25}
 \; u_{xxxxx}&=0,
\end{aligned}
\end{gather}
and
\begin{gather}
\label{KKK}
\begin{aligned}
u_{t} - 5 (u^{2} + w) u_{x} - 5 u w_{x} + 5 w_{y} + \epsilon^{2} (
u u_{xxx} +  w_{xxx} + \frac{5}{2} u_{xxx}) -\frac{
\epsilon^{4}}{25}
 \; u_{xxxxx}&=0,
\end{aligned}
\end{gather}
respectively \cite{Kon4}. Their dispersionless limits coincide:
\begin{equation}
u_{t} = 5 (u^{2} + w) u_{x} + 5 u u_{y} - 5 w_{y}. \label{dKK}
\end{equation}

\subsection{Symmetric case}

Here we list  known examples of the form (\ref{sym}).  The
relations $v_y=u_x$ and $w_x=u_y$ will be automatically assumed
whenever $v$ and $w$  appear explicitly in the equation. It is
quite remarkable that the `symmetric'  list is very restrictive,
and contains only two examples.

\bigskip

 \noindent{\bf Example 1.}
The  Veselov-Novikov  equation,
\begin{equation}
\label{vns} u_{t} =(uv)_x+(u w)_{y} +\epsilon^2  u_{xxx}+\epsilon
^2 u_{yyy},
\end{equation}
was introduced in \cite{VesNov}, \cite{Nizhnik}. It has the
dispersionless limit
\begin{equation}
\label{dvns} u_{t} =(uv)_x+(u w)_y.
\end{equation}

\bigskip

 \noindent{\bf Example 2.}
The  modified Veselov-Novikov  equation,
\begin{equation}
\label{vn} u_{t} =(uv)_x+(u w)_y +\epsilon^2  \left(
u_{xx}-\frac{3}{4}\frac{u_x^2}{u}\right)_x+ \epsilon^2  \left(
u_{yy}-\frac{3}{4}\frac{u_y^2}{u}\right)_y,
\end{equation}
was first introduced in \cite{Bogdanov} (in a somewhat different
form). It has the same dispersionless limit as in the previous
example,
\begin{equation}
\label{dvn} u_{t} =(uv)_x+(u w)_y.
\end{equation}

\bigskip

\section{Classification of integrable dispersionless limits}

In this section we classify integrable dispersionless equations of
the form (\ref{dnonsym}) and (\ref{dsym}) which may potentially
occur as dispersionless limits of integrable soliton equations of
the form (\ref{nonsym}) and (\ref{sym}), respectively.  The
integrability conditions are derived based on the method of
hydrodynamic reductions. For the convenience of the reader, we
briefly recall the main steps of this construction. As proposed in
\cite{Fer4}, the method of hydrodynamic reductions applies to
quasilinear equations of the following general form:
\begin{equation}
A ({\bf u}){\bf u}_t+B({\bf u}){\bf u}_x+C({\bf u}){\bf u}_y=0;
\label{quasi}
\end{equation}
here ${\bf u}=(u^1, ..., u^m)^t$ is an $m$-component column vector
of the dependent variables, and $A, B, C$ are $m\times m$
matrices. The method of hydrodynamic reductions consists of
seeking multi-phase solutions in the form
\begin{equation}
{\bf u}={\bf u}(R^1, ..., R^N) \label{phase}
\end{equation}
where the `phases' $R^i(x, y, t)$ are required to satisfy a pair
of consistent equations of hydrodynamic type,
$$
 R^i_y=\mu^i(R) R^i_x,  ~~~~ R^i_t=\lambda^i(R) R^i_x.
$$
We recall that the consistency
 conditions, $R^i_{yt}=R^i_{ty}$,
imply the following restrictions for the characteristic speeds
$\mu^i$ and $\lambda^i$:
$$
\frac{\partial_j\mu^i}{\mu^j-\mu^i}=\frac{\partial_j\lambda
^i}{\lambda^j-\lambda^i},
$$
$i\ne j, ~ \partial_i=\partial/\partial_{R^i}$, see \cite{Tsarev}.
The substitution of the ansatz (\ref{phase}) into  (\ref{quasi})
leads to a complicated over-determined system of PDEs for the
functions ${\bf u}(R), \ \mu^i(R)$ and $\lambda^i(R)$ whose
coefficients depend on the matrix elements of  $A, B, C$, and
their derivatives. In particular, the characteristic speeds $
\mu^i(R)$ and $\lambda^i(R)$  satisfy an algebraic relation ${\rm
det}(\lambda A+B+\mu C)=0$ which is nothing but the dispersion
relation of the system (\ref{quasi}). We will assume that the
dispersion relation  defines an irreducible algebraic curve of
degree $m$.


\medskip

\noindent {\bf Definition} \cite{Fer4}. {\it System (\ref{quasi})
is said to be {\it integrable} if, for any number of phases $N$,
it possesses infinitely many $N$-phase solutions  parametrized by
$2N$ arbitrary functions of one variable}.

\medskip

\noindent The requirement of the existence of such solutions
imposes strong constraints on the matrices $A, B, C$, which can be
effectively computed. Although these constraints are quite
formidable in general, there exists a  simple necessary condition
for the integrability which can be expressed in an invariant
differential geometric form as follows. Let us first introduce the
$m\times m$ matrix
$$
V=(\alpha A +\beta B+\gamma C)^{-1}(\tilde \alpha A +\tilde \beta
B+\tilde \gamma C)
$$
where $\alpha, \beta, \gamma$ and $\tilde \alpha, \tilde \beta,
\tilde \gamma $ are arbitrary constants. Given a $(1, 1)$-tensor
$V=[v^i_j]$, let us introduce the following objects:

\noindent Nijenhuis tensor
$$
{\cal
N}^i_{jk}=v^p_j\partial_{u^p}v^i_k-v^p_k\partial_{u^p}v^i_j-v^i_p(\partial_{u^j}v^p_k-\partial_{u^k}v^p_j),
$$
Haantjes tensor
$$
{\cal H}^i_{jk}={\cal N}^i_{pr}v^p_jv^r_k-{\cal
N}^p_{jr}v^i_pv^r_k-{\cal N}^p_{rk}v^i_pv^r_j+{\cal
N}^p_{jk}v^i_rv^r_p.
$$
One has the following result.

\begin{theorem} {\rm \cite{Fer8}} The vanishing of the Haantjes tensor is a necessary condition for the integrability of the system (\ref{quasi}).
\end{theorem}
Since  the Haantjes tensor can be obtained using computer algebra,
one gets an efficient integrability test (notice that all
components of the Haantjes tensor have to vanish for {\it any}
values of  the constants $\alpha, \beta, \gamma$ and $\tilde
\alpha, \tilde \beta, \tilde \gamma$). These necessary conditions
are very strong indeed, and in many cases turn out to be
sufficient. We point out that, for $m=2$, the Haantjes tensor
vanishes identically and does not produce any non-trivial
integrability conditions. In this case one proceeds as follows:
let us  multiply (\ref{quasi}) by $A^{-1}$, and diagonalize  $B$
(this is always possible in the $2$-component case). Thus, without
any loss of generality one can assume
$$
A= \left(\begin{array}{cc}
1 & 0  \\
0 & 1  \\
\end{array}
\right), ~~~ B= \left(\begin{array}{ccc}
a &  0 \\
0 & b  \\
\end{array}
\right), ~~~ C= \left(\begin{array}{ccc}
p & q  \\
r & s  \\
\end{array}
\right).
$$
In this particular normalization, the integrability conditions for
$2\times 2$ systems were obtained in \cite{Fer5}. These conditions
constitute a system of second order constraints for the
coefficients $a, b, p, q, r, s$ which can be easily tested. Let us
now apply this approach to the classification of integrable
systems of the form (\ref{dnonsym}) and (\ref{dsym}).

\subsection{Non-symmetric dispersionless limits}

Given an  equation of the form (\ref{dnonsym}),
$$
\begin{aligned}
u_t&=\varphi u_x+\psi u_y +\eta w_y, \\
 w_x&=u_y,
 \end{aligned}
$$
let us first  rewrite it in matrix form (\ref{quasi}) as follows:
$$
\left(\begin{array}{cc}
-1/\varphi & 0  \\
0 & 0  \\
\end{array}
\right) \left(\begin{array}{c}
u  \\
w
\end{array}
\right)_t+ \left(\begin{array}{cc}
1 & 0  \\
0 & 1  \\
\end{array}
\right) \left(\begin{array}{c}
u  \\
w
\end{array}
\right)_x+ \left(\begin{array}{cc}
\psi/\varphi & \eta/\varphi  \\
-1 & 0  \\
\end{array}
\right) \left(\begin{array}{c}
u  \\
w
\end{array}
\right)_y=0.
$$
This system is now in the form as studied in \cite{Fer5}. The
integrability conditions reduce to a system of second order
partial differential equations for the coefficients $\varphi,
\psi$ and $\eta$, which can be derived from the general
integrability conditions for $2\times2$ systems of hydrodynamic
type in $2+1$ dimensions as obtained in \cite{Fer5}:
\begin{gather}
\label{nonsymint}
\begin{aligned}
\varphi _{uu}&= -\frac{\varphi _w^2+\psi _u \varphi _w-2 \psi _w
\varphi _u}{\eta}, \\
\varphi _{uw}&= \frac{\eta _w \varphi _u}{\eta},\\
\varphi _{ww}&= \frac{\eta _w \varphi _w}{\eta },\\
\ \\
\psi _{uu}&= \frac{-\varphi _w \psi _w+\psi _u \psi _w-2
\varphi _w \eta _u+2 \eta _w \varphi _u}{\eta },\\
\psi _{uw}&= \frac{\eta _w \psi _u}{\eta },\\
\psi _{ww}&= \frac{\eta _w \psi _w}{\eta },\\
\ \\
\eta _{uu}&= -\frac{\eta _w \left(\varphi _w-\psi _u\right)}{\eta},\\
\eta _{uw}&= \frac{\eta _w \eta _u}{\eta },\\
\eta _{ww}&=\frac{\eta _w^2}{\eta };
\end{aligned}
\end{gather}
we assume $\eta \ne 0$: this is equivalent to the requirement that
the dispersion relation of the system (\ref{dnonsym}) defines an
irreducible conic (indeed, the condition ${\rm det}(\lambda
A+B+\mu C)=0$ is equivalent to $\lambda=\varphi+ \psi \mu +\eta
\mu^2$). We have verified that the system (\ref{nonsymint}) is in
involution, and all dispersionless limits appearing in Sect. 2.1
indeed satisfy these integrability conditions. Eqs.
(\ref{nonsymint}) are  straightforward to solve. First of all, the
equations for $\eta$ imply that, up to translations and
rescalings, $\eta=1$, $\eta=u$ or $\eta=e^wh(u)$. We will consider
all three possibilities case-by-case below. Notice  that $\varphi$
and $\psi$ are defined up to additive constants which can always
be set equal to zero via the Galilean transformations of the
initial equation (\ref{dnonsym}). Moreover, the system
(\ref{nonsymint}) is form-invariant under transformations of the
form
\begin{equation}
\tilde \varphi=\varphi -s\psi+s^2\eta, ~~~ \tilde \psi=
\psi-2s\eta, ~~~ \tilde \eta = \eta, ~~~  \tilde u = u, ~~~ \tilde
w=w+su, \label{s}
\end{equation}
$s$=const, which correspond to the following transformations
preserving the structure of equations  (\ref{dnonsym}):
 $$
 \tilde x=x-sy, ~~~ \tilde y=y, ~~~ \tilde u = u, ~~~ \tilde w=w+su.
 $$
 All our classification results are formulated modulo this  equivalence.

\medskip

\noindent {\bf Case 1: $\eta=1$}. Then the remaining equations
imply $\psi=\alpha w+f(u),\  \varphi=\beta w+g(u)$, where $f$ and
$g$ satisfy the linear  ODEs
$$
f''=\alpha(f'-\beta), ~~~ g''=2\alpha g'-\beta f'-\beta^2.
$$
The subcase $\alpha =0$ leads to polynomial solutions of the form
\begin{equation}
\psi=\gamma u, ~~~~
 \varphi=\beta w-\frac{1}{2}\beta (\beta+ \gamma)u^2+\delta u.
\label{pol}
\end{equation}
Up to equivalence transformations, the case $\alpha \ne 0$ leads
to exponential solutions,
\begin{equation}
\psi=\alpha w+\gamma e^{\alpha u}, ~~~~
 \varphi=\delta e^{2\alpha u};
\label{exp}
\end{equation}
 here $\alpha, \beta,  \gamma , \delta$ are arbitrary constants.

 \medskip

\noindent {\bf Case 2: $\eta=u$}. Then the remaining equations
imply $\psi=\alpha w+f(u),\  \varphi=\beta w+g(u)$, where $f$ and
$g$ satisfy the linear ODEs
$$
uf''=\alpha(f'-\beta)-2\beta, ~~~ ug''=2\alpha g'-\beta
f'-\beta^2.
$$
The case $\alpha \notin \{0,  -1, -1/2\}$ leads to power-like
solutions of the form
\begin{equation}
\psi=\alpha w+\gamma u^{\alpha +1}, ~~~~
 \varphi=\delta u^{2\alpha +1}.
\label{power}
\end{equation}
The subcase $\alpha =0$ leads  to logarithmic solutions,
\begin{equation}
\psi=-2\beta u\ln u -\beta   u, ~~~~
 \varphi=\beta w+\beta^2u\ln^2 u+\delta u.
\label{log1}
\end{equation}
The subcase $\alpha =-1$ gives
\begin{equation}
\psi =-w+\gamma \ln u, ~~~~ \varphi=\delta/u. \label{log2}
\end{equation}
Finally, the subcase $\alpha =-1/2$ gives
\begin{equation}
\psi =-\frac{1}{2}w+\gamma \sqrt u, ~~~~ \varphi=\delta \ln u.
\label{log3}
\end{equation}

\medskip

\noindent {\bf Case 3: $\eta=e^wh(u)$}. Then the remaining
equations imply $\psi=e^ wf(u),\  \varphi=e^ wg(u)$ where $f$, $g$
and $h$ satisfy the nonlinear system of ODEs
$$
h''=f'-g, ~~~~ g''h=2fg'-gf'-g^2, ~~~~ f''h=2hg'-2gh'+ff'-fg.
$$
 Setting $g=p', \ f=h'+p$, we can rewrite this system as a pair of third order ODEs
 $$
 hp'''=2h'p''-p'h''+2pp''-2p'^2, ~~~ hh'''=h'h''-2h'p'+hp''+ph'',
 $$
 which, up to a change of sign $p\to -p$, identically coincides with a system arising  in the classification of integrable conservative hydrodynamic chains (subcase $I_1$ of Sect. 3.1 in \cite{Mar}).
Setting  $p=h'$, the second equation will be satisfied
identically, while  the first  one implies a fourth order ODE for
$h$,
 $h''''h+3(h'')^2-4h'h'''=0$, whose general solution is an elliptic sigma-function: $h=\sigma (u)$, here $(\ln \sigma)''=-\wp, \  (\wp')^2=4\wp^3-c$
(notice that $g_2=0, \ g_3=c$). Thus,  as a particular case we
have
$$
h=\sigma(u), ~~~ f=2\sigma'(u), ~~~ g=\sigma ''(u).
$$
Another subclass of solutions can be obtained by setting $p=ch$
which implies
$$
h'''h-h''h'=2c(h''h-h'^2)
$$
with the general solution
$$
h=\alpha e^{(c+\gamma)u}+\beta e^{(c-\gamma)u};
$$
here $ \alpha, \beta, \gamma$ are arbitrary constants. Although
the structure of the general solution  is quite complicated, one
can show that Case 3 cannot arise as a  dispersionless limit of an
integrable third order soliton equation.

\subsection{Symmetric dispersionless limits}

In this section we consider first order equations of the form
(\ref{dsym}),
$$
\begin{aligned}
u_t&=\varphi u_x+\psi u_y +\eta w_y+\tau v_x, \\
 w_x&=u_y, \\
  v_y&=u_x,
\end{aligned}
$$
where the coefficients $\varphi, \psi, \eta, \tau$ are functions
of $u, v, w$. We assume that the dispersion relation of this
system defines an irreducible cubic, which is equivalent to the
requirement  $\eta\ne 0$ and $\tau \ne 0$ (indeed, the dispersion
relation has the form $\lambda \mu =\tau + \varphi \mu + \psi
\mu^2 +\eta \mu^3$). In this case the integrability conditions
reduce to a  system of first order partial differential equations
for the coefficients $\varphi, \psi, \eta$ and $\tau$ which can be
obtained from  the requirement of the vanishing of the Haantjes
tensor \cite{Fer8} as outlined in Sect.\ 3. The details are as
follows: first we rewrite Eq. (\ref{dsym}) in matrix form,
$$
A{\bf {u}}_t+B{\bf {u}}_x+C{\bf {u}}_y=0,
$$
where ${\bf u}$ is a three-component column vector ${\bf u}=(u, v,
w)^t$, and $A, B, C$ are  $3\times 3$ matrices,
$$
A= \left(\begin{array}{ccc}
-1 & 0 & 0 \\
0 & 0 & 0 \\
0 &0 & 0
\end{array}
\right), ~~~ B= \left(\begin{array}{ccc}
\varphi & \tau & 0 \\
0 & 0 & 1 \\
-1 &0 & 0
\end{array}
\right), ~~~ C= \left(\begin{array}{ccc}
\psi & 0 & \eta \\
-1 & 0 & 0 \\
0 &1 & 0
\end{array}
\right).
$$
The necessary conditions for integrability can be obtained from
the requirement of the vanishing of the Haantjes tensor of the
following family of matrices,
$$
(\alpha A +\beta B+\gamma C)^{-1}(\tilde \alpha A +\tilde \beta
B+\tilde \gamma C).
$$
In fact, it is sufficient to require the vanishing of the Haantjes
tensor for a two-parameter family $(\alpha A + B)^{-1}(\tilde
\alpha A +C)$. This condition
 turns out
to be very restrictive, and leads to the following constraints for
the coefficients  $\varphi, \psi, \eta$ and $\tau$:
\begin{align*}
&\tau_{u} = \varphi_{v},& \qquad  &\eta_{u} =
\psi_{w},&  \\
&\tau_{v} = \frac{\tau}{\eta} \psi_{u},&
\qquad &\eta_{v} = 0,&   \\
 &\tau_{w} = 0, &  \qquad &\eta_{w} = \frac{\eta}{\tau} \varphi_{u},& \\
 & \psi_{v} = \varphi_w=0, &  \qquad & \tau \psi_{w} = {\eta} \varphi_{v}.
 \end{align*}
 The integration of this system is straightforward. First of all, one can set
 $\psi =f_u, ~ \eta =f_w$ and $ \varphi=g_u, ~ \tau = g_v$ where $f=f(u, w)$ and $g=g(u, v)$. The separation of variables leads to the relations
 \begin{align*}
&f_{w} = a(w)k(u),& \qquad  &g_v=b(v)k(u), &  \\
&f_{uu} =\beta a(w)k(u),& \qquad &g_{uu}=\alpha b(v)k(u),
 \end{align*}
where the functions $a(w), \ b(v)$ and $k(u)$ satisfy the ODEs
$a'=\alpha a, \ b'=\beta b$ and $k''=\alpha\beta k$; here $\alpha$
and $\beta$ are arbitrary constants. Up to elementary
translations, rescalings and Galilean transformations, this leads
to the following subcases:

 \medskip

 \noindent {\bf Case 1.} $\alpha=\beta =0$. This leads to equations of the form
 $$
 u_t=\nu (uv)_x+\mu (uw)_y,
 $$
where $\mu, \nu$ are arbitrary constants. These correspond to the
Veselov-Novikov cases from Sect. 2.2.

 \medskip

 \noindent {\bf Case 2.} $\alpha \ne 0, \ \beta =0$. This leads to equations of the form
 $$
 u_t=\nu (uv+\alpha u^3/6)_x+\mu (e^{\alpha w}u)_y,
 $$
and
$$
 u_t=\nu (v+\alpha u^2/2)_x+\mu (e^{\alpha w})_y,
 $$
here $\mu, \nu, \alpha $ are arbitrary constants.

\medskip

 \noindent {\bf Case 3.} $\alpha \ne 0, \ \beta \ne 0. $ This leads to equations of the form
 $$
 u_t=\nu (e^{\beta v}k(u))_x+ \mu (e^{\alpha w}k(u))_y,
 $$
where $\nu, \mu, \alpha, \beta$ are arbitrary constants, and
$k''=\alpha \beta k$.

\section{Classification of integrable 3rd order dispersive equations}

Given an integrable dispersionless limit, one has to reconstruct
dispersive terms. This can be done by requiring that all
hydrodynamic reductions of the dipersionless system are inherited
by its dispersive counterpart. We will  illustrate this procedure
using the  KP equation,
$$
u_{t} = u u_{x} +w_y+ \epsilon^{2} u_{xxx}, ~~~~ w_x=u_y.
$$
Its dispersionless limit, the dKP equation,
$$
u_{t} = u u_{x}+w_y, ~~~~ w_x=u_y,
$$
possesses one-phase solutions of the form  $u=R$, $w=w(R)$ where
the phase $R(x, y, t)$ satisfies a pair of Hopf-type equations
\begin{gather}
\label{R}
\begin{aligned}
R_{y} = \mu R_{x} , ~~~~ R_{t} =(\mu^{2} + R) R_{x};
\end{aligned}
\end{gather}
here $\mu(R)$ is an arbitrary function, and $w'=\mu$.
Equivalently, one can say that Eqs. (\ref{R}) constitute a
one-component hydrodynamic reduction of the dKP equation. Although
the dKP equation is known to possess infinitely many $N$-component
reductions for arbitrary $N$ \cite{Gibb94, GibTsa96, GibTsa99,
Kodama},  one-component reductions will be sufficient for our
purposes. The main observation of \cite{FerM} is that {\it all}
one-component reductions (\ref{R})  can be deformed into
reductions of the full KP equation by adding appropriate
dispersive terms which are {\it polynomial} in the $x$-derivatives
of $R$. Explicitly, one has the following formulae for the
deformed one-phase solutions,
\begin{equation}
u=R, ~~~ w=w(R)+\epsilon^{2}\left(\mu' R_{xx} +\frac{1}{2}(\mu''-
(\mu')^3) R_{x}^2 \right)+ O(\epsilon^{4}), \label{uw_Def}
\end{equation}
notice that one can always assume that $u$ remains undeformed
modulo the Miura group \cite{Dub1}. The deformed equations
(\ref{R}) take the form
\begin{gather}
\label{R_Def}
\begin{aligned}
R_{y} =& \mu R_{x} \\
& +\epsilon^{2} \left(\mu' R_{xx} +\frac{1}{2}
(\mu''- (\mu')^3) R_{x}^2 \right)_x + O(\epsilon^{4}),\\
R_{t} =& (\mu^{2} + R) R_{x} \\
&+\epsilon^{2} \left( (2\mu \mu'+1)R_{xx}+( \mu \mu''-\mu
(\mu')^3+(\mu')^2/2) R_{x}^2 \right)_x  + O(\epsilon^{4}).
\end{aligned}
\end{gather}
In other words, the KP equation can be `decoupled' into a pair of
$(1+1)$-dimensional equations (\ref{R_Def}) in  infinitely many
ways, indeed, $\mu(R)$ is an arbitrary function. The series in
(\ref{uw_Def}) and (\ref{R_Def}) contain only even powers of
$\epsilon$, and do not terminate in general.

Conversely, the requirement of the inheritance of all
one-component reductions allows one to reconstruct  dispersive
terms:  given the dKP equation, let us look for a third order
dispersive extension in the form
\begin{equation}
u_{t} = u u_{x}+w_y+\epsilon(...)+\epsilon^2(...), ~~~~ w_x=u_y,
\label{3.1}
\end{equation}
where the terms at $\epsilon$ and $\epsilon^2$ are homogeneous
differential polynomials in the $x$- and  $y$-derivatives of $u$
and $w$ of the order two and three, respectively, whose
coefficients are allowed to be arbitrary functions of $u$ and $w$.
We require that all one-component reductions (\ref{R}) can be
deformed accordingly, so that we have the following analogues of
Eqs. (\ref{uw_Def}) and (\ref{R_Def}),
\begin{equation}
u=R, ~~~ w=w(R)+\epsilon (...)+ \epsilon^{2}(...)+
O(\epsilon^{3}), \label{3.2}
\end{equation}
and
\begin{equation}
R_{y} = \mu R_{x} +\epsilon (...)+
\epsilon^2(...)+O(\epsilon^{3}), ~~~~ R_{t} =(\mu^{2} + R)
R_{x}+\epsilon (...)+\epsilon^2(...)+O(\epsilon^{3}), \label{3.3}
\end{equation}
respectively.  In Eqs. (\ref{3.2}) and (\ref{3.3}), dots denote
terms which are polynomial in the derivatives of $R$. Substituting
Eqs. (\ref{3.2}) into (\ref{3.1}), and using (\ref{3.3}) along
with the consistency conditions $R_{ty}=R_{yt}$, one arrives at a
complicated set of relations allowing one to uniquely reconstruct
dispersive terms in (\ref{3.1}): not surprisingly, we obtain that
all terms at $\epsilon $ vanish, while the terms at $\epsilon^2$
result in the familiar KP equation. Moreover, one only needs to
perform calculations up to the order $\epsilon^4$  to arrive at
this result! It is important to emphasize that the above procedure
is required to work for {\it arbitrary } $\mu$: whenever one
obtains a differential polynomial in $\mu$ which has to vanish due
to the  consistency conditions, all its coefficients have to be
set equal to zero independently. Another observation is that the
reconstruction procedure does not necessarily lead to a unique
dispersive extension as in the dKP case: one and the same
dispersionless system may possess essentially non-equivalent
dispersive extensions. In most of the cases one can get the
necessary classification results working with one-component
reductions only. There is however one particular situation where
one-component reductions are not sufficient. This is explained in
the remark below.

\noindent {\bf Remark 1.} Let us consider the dKP equation,
$$
u_{t} = u u_{x}+w_y, ~~~~ w_x=u_y;
$$
its one-component reductions (\ref{R}) can be shown to satisfy a
pair of additional first order constraints,
$$
u_y^2-u_xw_y=0, ~~~~~ (w_t-uu_y)u_x-u_yw_y=0.
$$
Conversely, any solution satisfying these constraints comes from
one-component reductions. Similarly, one can show that
two-component reductions of dKP are characterised by  a pair of
second order differential constraints, etc. Let us introduce an
extension of dKP in the form
$$
u_{t} = u u_{x}+w_y+\ep(u_y^2-u_xw_y), ~~~~ w_x=u_y;
$$
by construction, it inherits all {\it undeformed} one-component
reductions: the $\ep$-term vanishes on one-component reductions
identically. This extension is, however, not integrable: one can
show that it is not consistent with the requirement of the
inheritance of $N$-component reductions for $N\geq 2$. Thus, in
what follows we eliminate deformations which inherit undeformed
one-component reductions.

\medskip

In general, we proceed as follows.  For definiteness, we will
outline the algorithm for integrable dispersionless equations of
the form (\ref{dnonsym}),
$$
\begin{aligned}
u_t&=\varphi u_x+\psi u_y +\eta w_y, ~~~~
 w_x=u_y.
 \end{aligned}
$$
Its one-component reductions are of the form $u=R$, $w=w(R)$ where
$R(x, y, t)$ satisfies a pair of Hopf-type equations
\begin{gather*}
\begin{aligned}
R_{y} = \mu R_{x} , ~~~~ R_{t} =(\varphi +\psi \mu+ \eta \mu^{2})
R_{x};
\end{aligned}
\end{gather*}
here $\mu(R)$ is an arbitrary function, and $w'=\mu$. We seek a
third order dispersive deformation of  Eq. (\ref{dnonsym}) in the
form
$$
\begin{aligned}
u_t&=\varphi u_x+\psi u_y +\eta w_y+\epsilon
(...)+\epsilon^{2}(...), ~~~~
 w_x=u_y,
 \end{aligned}
$$
and postulate that one-phase solutions can be deformed
accordingly,
$$
u=R, ~~~ w=w(R)+\epsilon(...)+\epsilon^{2}(...)+ O(\epsilon^{3}),
$$
where
$$
R_{y} = \mu R_{x} +\epsilon (...)+\epsilon^2(...)+O(\epsilon^{3}),
~~~~ R_{t} =(\varphi +\psi \mu+ \eta \mu^{2})
R_{x}+\epsilon(...)+\epsilon^2(...)+O(\epsilon^{3}).
$$
Proceeding as outlined above we reconstruct possible dispersive
terms. In fact, one can start with arbitrary $\varphi, \psi,
\eta$: our procedure will eventually recover the constraints
obtained in Sect. 3. However, using the classification results of
Sect. 3 from the very beginning considerably simplifies the
calculations.

\noindent {\bf Remark 2.} We point out that the formulae for
dispersive deformations contain the expression
$$
\eta_w \mu^3+(\psi_w+\eta_u)\mu^2+(\varphi_w+\psi_u)\mu+\varphi_u
$$
in the denominator. Since $\mu$ is assumed to be arbitrary, this
expression is nonzero unless $\varphi, \psi, \eta$ satisfy the
relations
\begin{equation}
\eta_w=0, ~~~ \psi_w+\eta_u=0, ~~~ \varphi_w+\psi_u=0, ~~~
\varphi_u=0. \label{lindeg}
\end{equation}
These relations characterize the so-called {\it totally linearly
degenerate systems},  which are known to be quite special from the
point of view of the global existence of classical solutions: it
was conjectured in  \cite{Majda} that smooth initial data for
totally linearly degenerate systems do not break down in finite
time. Modulo the integrability conditions (\ref{nonsymint}), the
relations (\ref{lindeg}) lead to equations of the form
$$
u_t=\alpha (wu_x-uw_x)+\beta(wu_y-uw_y)+\gamma w_y, ~~~ w_x=u_y,
$$
which have been discussed before in the context of the so-called
`universal hierarchy' \cite{Shabat}. For totally linearly
degenerate systems (in particular, for linear systems), the
procedure based on deformations of hydrodynamic reductions does
not work, as the following simple example shows. Let us consider
the KP equation,
$$
u_{t} =\alpha u u_{x} +w_y+ \epsilon^{2} u_{xxx}, ~~~~ w_x=u_y,
$$
where we introduced a parameter $\alpha$: for $\alpha =0$ the
equation becomes linear. Looking for deformed one-phase solutions
in the form
$$
u=R, ~~~ w=w(R)+ \epsilon^{2}(...)+ O(\epsilon^{4}),
$$
where
$$
R_{y} = \mu R_{x} + \epsilon^2(...)+O(\epsilon^{4}), ~~~~ R_{t}
=(\mu^{2} +\alpha R) R_{x}+\epsilon^2(...)+O(\epsilon^{4}),
$$
one can obtain the relation $\alpha b(R)-\mu'=0$ where $b(R)$ is
the coefficient at $R_{xxx}$ in the $\epsilon^2$-term in the
expansion of $R_y$. For $\alpha =0$ one cannot solve for $b(R)$,
and obtains a relation $\mu'=0$. Thus, the linear equation $u_{t}
=w_y+ \epsilon^{2} u_{xxx}$ does not inherit generic hydrodynamic
reductions of its dispersionless limit. Another example of this
kind is provided by the potential KP equation,
\begin{equation}
\label{vn} u_{t} = w_y +\frac{\epsilon}{2}u_x^2+\epsilon^2
u_{xxx}.
\end{equation}
One can show that this equation does not inherit  hydrodynamic
reductions of its dispersionless limit. However, some particular
reductions can be inherited, for instance, those with $\mu$=const.

Thus, we  exclude  totally linearly degenerate systems from the
further considerations: dispersive deformations of such systems do
not inherit hydrodynamic reductions, and require a different
approach.

\bigskip

\subsection{Non-symmetric dispersive equations}

In this Section we summarize the classification results  for
integrable non-symmetric third order  equations (\ref{nonsym}),
$$
u_t=\varphi u_x+\psi u_y +\eta w_y+\epsilon(...)+\epsilon^2(...),
~~~~ w_x=u_y,
$$
which are obtained by adding dispersive terms to integrable
dispersionless candidates from Sect. 3.1.  Thus, we follow the
classification of Sect. 3.1.

\medskip

\noindent {\bf Case 1:} We have verified that the exponential
solutions (\ref{exp}) do not survive, so that all non-trivial
examples come from the polynomial case (\ref{pol}),
$$
\eta=1, ~~~~ \psi=\gamma u, ~~~~
 \varphi=\beta w-\frac{1}{2}\beta (\beta+ \gamma)u^2+\delta u.
$$
We point out that the corresponding dispersionless system
possesses the Lax pair
\begin{gather}
\label{lax0}
\begin{aligned}
S_y &= \beta u S_x+r(S_x),  \\
S_t &= \left(\beta w +\frac{1}{2}\beta(\beta + \gamma) u^2\right)
S_x+\beta u S_xr'(S_x)+z(S_x),
\end{aligned}
\end{gather}
where
$$
r(S_x)=-\frac{\delta}{\beta + \gamma}S_x+S_x^{\frac{2\beta
+\gamma}{\beta}}, ~~~ z'=r'^2.
$$
Lax pairs of this kind, consisting of two compatible
Hamilton-Jacobi type equations, were first introduced by Zakharov
in \cite{Zakharov}. A detailed  analysis of dispersive
deformations leads to the two branches: $\gamma=0$, which
corresponds to the $(2+1)$-dimensional Gardner equation (Example 3
of Sect. 2.1), and the case $\gamma =-3\beta$. In the latter case
one can set $\delta=0$, which leads to  the apparently new
equation (\ref{new1}),
$$
u_t=(\beta w+\beta^2u^2)u_x-3\beta u u_y+w_y+\epsilon^2
[B^3(u)-\beta B^2(u)u_x],
$$
where $B=\beta uD_x- D_y$. The dispersionless limit of this
equation possesses the Lax pair
\begin{gather}
\label{lax}
\begin{aligned}
S_xS_y &= \beta u S_x^2+\frac{1}{3},  \\
S_t &= {\beta^3}u^3S_x^3-  S_y^3+\beta wS_x,
\end{aligned}
\end{gather}
which follows from (\ref{lax0}) when $\gamma =-3\beta$. Its
dispersive extension  is
\begin{gather}
\label{laxdisp}
\begin{aligned}
\psi_{xy} &= \beta u \psi_{xx}+\frac{1}{3\epsilon^2} \psi ,  \\
\psi_t &=\beta^3\epsilon^2u^3\psi_{xxx} -
\epsilon^2\psi_{yyy}+3\beta^2\epsilon^2uu_y\psi_{xx}+\beta
w\psi_x.
\end{aligned}
\end{gather}
This is case (\ref{new1}) from the Introduction.

\medskip

\noindent {\bf Case 2:} One can prove that none of the logarithmic
cases (\ref{log1}), (\ref{log2})  and (\ref{log3})  survive, so
that all non-trivial examples come from the power case
(\ref{power}),
$$
\eta=u, ~~~~ \psi=\alpha w+\gamma u^{\alpha +1}, ~~~~
 \varphi=\delta u^{2\alpha +1}.
$$
Further analysis leads to the following branches.

\noindent {\small{\bf Subcase 2.1:}}  $\alpha =1$. In this case
$$
\eta=u, ~~~~ \psi= w+\gamma u^{2}, ~~~~
 \varphi=\delta u^{3}.
$$
The corresponding dispersionless Lax pair is of the form
\begin{gather}
\label{laxalpha=1}
\begin{aligned}
S_y &=ua ,  \\
S_t &= uw a+\frac{1}{3}a(\gamma +a')u^3,
\end{aligned}
\end{gather}
where the function $a(S_x)$  solves the ODE
$aa''-2a'^2=3\delta+2\gamma a'.$ The further analysis gives either
$\gamma=\delta=0$, which leads to the non-symmetric
Veselov-Novikov cases (Examples 4 and 5 of Sect. 2.1, in this case
one can take $a=1/S_x$), or $\delta=\frac{4}{27}\gamma^2$, in
which case one arrives at the apparently new dispersive equation
(\ref{new2}),
$$
u_t=\frac{4}{27}\gamma^2 u^3u_x+(w+\gamma u^2) u_y+uw_y+\ep^2
[B^3(u)-\frac{1}{3}\gamma u_xB^2(u)],
$$
where $B=\frac{1}{3}\gamma u D_x+D_y$. This corresponds to the
choice $a=1/S_x-\frac{\gamma}{3}S_x$ in the dispersionless Lax
pair (\ref{laxalpha=1}), which gives
\begin{gather}
\label{lax11}
\begin{aligned}
S_xS_y &= -\frac{\gamma}{3} u S_x^2-\frac{u}{3},  \\
S_t &= \frac{\gamma^3}{27}u^3S_x^3+
S_y^3+\frac{\gamma^2}{27}u^3S_x+ wS_y.
\end{aligned}
\end{gather}
The dispersive extension of this Lax pair is
\begin{gather}
\label{lax11disp}
\begin{aligned}
\psi_{xy} &=-\frac{\gamma}{3} u \psi_{xx}-\frac{1}{3\epsilon^2}u\psi ,  \\
\psi_t &=\frac{\epsilon^2\gamma^3}{27}u^3\psi_{xxx}
+\epsilon^2\psi_{yyy}-\frac{\epsilon^2\gamma^2}{3}uu_y\psi_{xx}+\frac{\gamma^2}{27}u^3\psi_x+w\psi_y-\frac{\gamma}{3}uu_y\psi.
\end{aligned}
\end{gather}
The transformation $\gamma \to 3\beta, \ y\to -y, \ w\to -w$
reduces this case to Eq. (\ref{new2}) from the Introduction.

\noindent {\small{\bf Subcase 2.2:}}  $\alpha =-2$. In this case
one obtains $\gamma=0$, while $\delta$ can be an arbitrary
constant. The corresponding dispersive extension takes the form
(\ref{new3}),
$$
 u_{t} =\frac{\delta}{u^3}u_x-2w  u_{y} +u w_y -\frac{\epsilon^2}{u}\left(\frac{1}{u}\right)_{xxx},
$$
for $\delta=0$ it reduces to the Harry Dym equation (Example 6 of
Sect. 2.1). The dispersionless limit of this equation possesses
the Lax pair
\begin{gather}
\label{laxHD}
\begin{aligned}
S_y &=\frac{S_x^2+\tau}{u^2},  \\
S_t &= -2w\frac{S_x^2+\tau}{u^2}+ \frac{4}{3}\frac{S_x^3+\tau
S_x}{u^3};
\end{aligned}
\end{gather}
here $\tau=3\delta/4$. Its dispersive extension is of the form
$L_t=[A, L]$  where
\begin{gather}
\label{laxdelta}
\begin{aligned}
L &= \frac{\ep^2}{u^2} D_x^2+\frac{\ep}{\sqrt 3} D_y+\frac{\delta}{4u^2},  \\
A &=\frac{4\ep^2}{u^3 } D_x^3+\left(-\frac{6\ep^2
u_x}{u^4}+\frac{2\sqrt 3 \ep w}{u^2}\right)
D_x^2+\frac{\delta}{u^3} D_x+\left(-\frac{3\delta
u_x}{2u^4}+\frac{\sqrt 3 \delta w}{2\ep u^2}\right).
\end{aligned}
\end{gather}

\medskip

\noindent {\bf Case 3:} One can show that none of the examples
from this class possess  third order dispersive extensions.

\subsection{Symmetric dispersive equations}

A detailed analysis  of dispersive extensions of the form
(\ref{sym}),
$$
u_t=\varphi u_x+\psi u_y +\eta w_y+\tau
v_x+\epsilon(...)+\epsilon^2(...), ~~~~ w_x=u_y,~ v_y=u_x,
$$
does not give any new examples: everything reduces to the two
cases of Sect. 2.2. Notice that both symmetric VN and mVN
equations can be viewed as linear combinations of the two
commuting non-symmetric counterparts thereof.

\section{Concluding remarks}

We have proposed a new approach to the classification of
integrable equations in $2+1$ dimensions based on the concept of
hydrodynamic reductions and their dispersive deformations. It
consists of the two steps:

\noindent --- Classification of dispersionless systems which may
(potentially) arise as dispersionless limits of soliton equations.
This can be efficiently achieved using the method of hydrodynamic
reductions as outlined in \cite{Fer4};

\noindent --- Classification of possible dispersive deformations
based on the requirement that hydrodynamic reductions of the
dispersionless limit are inherited by the dispersive equation
\cite{FerM}.

\noindent This procedure was applied to the classification of
third order soliton equations with `simplest'  nonlocalities.
Further research in this direction may include the following
topics:

\noindent (a) Classification of more general (in particular,
higher order) soliton equations/systems with more complicated
structure of nonlocal terms. Thus, one may allow `nested'
nonlocalities of the type $w=D_x^{-1}D_yu$,  $v=D_x^{-1}D_yF(u,
w)$, etc.

\noindent (b) Construction of dispersive deformations via an
appropriate quantization of the corresponding dispersionless Lax
pairs \cite{Zakharov}.

\noindent (c) Investigation of the structure of multi-soliton
solutions of the new equations (\ref{new1}) -- (\ref{new3})
 in the spirit of \cite{Chak1, Chak2}.

\section*{Acknowledgements}

We thank B. Dubrovin,   A. Hone,  B. Konopelchenko, A. Mikhailov,
M. Pavlov  and J.P. Wang for clarifying discussions. We also thank
the referees for useful comments. The research of EVF and AM  was
supported by the EPSRC grant EP/D036178/1,  the European Union
through the FP6 Marie Curie RTN project ENIGMA (Contract number
MRTN-CT-2004-5652), and the ESF programme MISGAM. The research of
VSN was supported by the EPSRC Postdoctoral Fellowship grant
EP/C527747/1.


\begin{thebibliography}{99}

\bibitem{Baikov} V.A. Baikov, R.K Gazizov and  N.Kh. Ibragimov, Approximate symmetries and formal linearization, J. Appl. Mech. Tech. Phys. {\bf 30}, no. 2 (1989) 204--212.






\bibitem{Bogdanov} L.V. Bogdanov, Veselov-Novikov equation as a natural
two-dimensional generalization of the Korteweg-de Vries equation,
Theor. and Math. Phys. {\bf 70} (1987) 309--314.

\bibitem{Bog} O.I. Bogoyavlenskii, Overturning solitons in two-dimensional integrable equations, Russian Math. Surveys {\bf 45}, no. 4 (1990) 1--86.

\bibitem{Boiti} M. Boiti, J. Jp. Leon, M. Manna, F. Pempinelli, On the spectral transform of a Korteweg de Vries equation in two spatial dimensions,  Inverse Problems {\bf 2}, no. 3 (1986) 271--279.

\bibitem{Chak1} S. Chakravarty and Y Kodama, Classification of the line-soliton solutions of KPII,  J. Phys. A {\bf 41}, no. 27 (2008) 275209, 33 pp.

\bibitem{Chak2}  S. Chakravarty and  Y. Kodama,
Soliton solutions of the KP equation and application to shallow
water waves, arXiv:0902.4433.



\bibitem{Dub1} B.A. Dubrovin and Youjin  Zhang, Bi-Hamiltonian hierarchies in $2$D
topological field theory at one-loop approximation,  Comm. Math.
Phys. {\bf 198} (1998) no. 2, 311--361.


\bibitem{Dub2} B.A. Dubrovin, Si-Qi Liu and  Youjin Zhang,  On Hamiltonian perturbations of hyperbolic systems of conservation laws. I. Quasi-triviality of bi-Hamiltonian perturbations,  Comm. Pure Appl. Math. {\bf 59}, no. 4 (2006) 559--615.

\bibitem{Dub3} B.A. Dubrovin,  On Hamiltonian perturbations of hyperbolic systems of conservation laws. II. Universality of critical behaviour, Comm. Math. Phys. {\bf 267}, no. 1 (2006) 117--139.



\bibitem{Fer4} E.V. Ferapontov and K.R. Khusnutdinova, On integrability of
(2+1)-dimensional quasilinear systems, Comm. Math. Phys.  {\bf
248} (2004) 187--206.


\bibitem{Fer5} E.V. Ferapontov and K.R. Khusnutdinova, The characterization of 2-component (2+1)-dimensional integrable systems of hydrodynamic type, J. Phys. A: Math. Gen. {\bf 37}, no. 8 (2004)
2949--2963.



\bibitem{Fer8} E.V. Ferapontov and K.R. Khusnutdinova, Double waves in multi-dimensional systems of hydrodynamic type: the necessary condition for integrability,  Proc. Royal Soc.  A  {\bf 462} (2006) 1197--1219.


\bibitem{Mar} E.V. Ferapontov and D.G. Marshall, Differential-geometric approach to the integrability of hydrodynamic chains: the Haantjes tensor,  Math. Ann. {\bf 339}, no. 1 (2007)  61--99.


\bibitem{FerM} E.V. Ferapontov and A. Moro,
Dispersive deformations of hydrodynamic reductions of 2D
dispersionless integrable systems, J. Phys. A: Math. Theor. 42
(2009) 035211, 15pp.







\bibitem{Gibb94} J. Gibbons and Y. Kodama,   A method for solving the
dispersionless KP hierarchy and its exact solutions. II, Phys.
Lett. A {\bf135} (1989) 167--170.

\bibitem{GibTsa96} J. Gibbons and S.P. Tsarev,
Reductions of the Benney equations, Phys. Lett. A {\bf 211} (1996)
19--24.

\bibitem{GibTsa99} J. Gibbons and S.P. Tsarev, Conformal maps and
reductions of the Benney equations, Phys. Lett. A {\bf 258} (1999)
263--271.

\bibitem{Kodama} Yu. Kodama, A method for solving the dispersionless KP equation and its exact
solutions, Phys. Lett. A {\bf 129}, no. 4 (1988) 223--226.

\bibitem{KodMik} Yu. Kodama and A.V. Mikhailov,  Obstacles to asymptotic integrability. Algebraic aspects of integrable systems, 173--204, Progr. Nonlinear Differential Equations Appl., 26, Birkh?user Boston, Boston, MA, 1997.




\bibitem{Kon4} B.G. Konopelchenko and V.G. Dubrovsky,  Some new integrable nonlinear evolution equations in 2+1 dimensions,  Phys. Letters A {\bf 102}, N 1, 2 (1984) 15--17.



\bibitem{Zhang} Si-Qi Liu and Youjin Zhang,  On quasi-triviality and integrability of a class of scalar evolutionary PDEs, J. Geom. Phys. {\bf 57}, no. 1 (2006) 101--119.


\bibitem{Majda} A. Majda, Compressible fluid flow and systems of conservation laws in several space variables, Applied Mathematical Sciences, 53, Springer-Verlag, New York (1984) 159 pp.


\bibitem{Mik0} A.V. Mikhailov, A. B Shabat and V.V. Sokolov,  The symmetry approach to classification of integrable equations. What is integrability?, 115--184, Springer Ser. Nonlinear Dynam., Springer, Berlin, 1991.



\bibitem{Mik1} A.V.  Mikhailov and R.I. Yamilov, Towards classification of $(2+1)$-dimensional integrable equations. Integrability conditions, I. J. Phys. A {\bf 31}, no. 31 (1998) 6707--6715.

\bibitem{Mik2} A.V. Mikhailov and V.S.  Novikov,  Perturbative symmetry approach, J. Phys. A {\bf 35}, no. 22 (2002) 4775--4790.




\bibitem{Shabat} L. Martinez Alonso and A.B. Shabat, Hydrodynamic reductions and solutions of a universal hierarchy, Teoret. Mat. Fiz. {\bf 140} (2004), 216--229.

\bibitem{Nizhnik} L.P. Nizhnik, Integration of multidimensional nonlinear
equations by the method of inverse problem, DAN SSSR, {\bf 254}
(1980) 332.

\bibitem{Sakovich} S.Yu. Sakovich,  Fujimoto-Watanabe equations and differential substitutions, J. Phys. A {\bf 24}, no. 10 (1991) L519--L521.



\bibitem{Sz} B.M. Szablikowski and M. Blaszak, Dispersionful analogue of the Whitham hierarchy, arXiv:0707.1082.



\bibitem{Tsarev} S.P. Tsarev, Geometry of Hamiltonian systems of hydrodynamic type. Generalized hodograph method, Izvestija AN USSR Math. {\bf 54}  (1990) 1048--1068.

\bibitem{Ustinov} N.V. Ustinov, Darboux transformations, infinitesimal symmetries
and conservation laws for the nonlocal two-dimensional Toda
lattice, J. Phys. A: Math. Gen. {\bf 35} (2002) 6963?6972.

\bibitem{VesNov} A.P. Veselov  and S.P. Novikov, Finite-gap two-dimensional
potential Schr\"odinger operators. Explicit formulae and evolution
equations, DAN SSSR, {\bf 279} (1984) 20.

\bibitem{Wang} Wang, Jing Ping, On the structure of $(2+1)$-dimensional commutative and noncommutative integrable equations. J. Math. Phys. {\bf 47}, no. 11 (2006) 113508, 19 pp.

\bibitem{watanabe} A. Fujimoto and  Y. Watanabe, Polynomial evolution equations of not normal type admitting nontrivial symmetries. Phys. Lett. A {\bf 136}, no. 6 (1989) 294--299.

\bibitem{MSS} A.V. Mikhailov, V.V. Sokolov and A.B. Shabat, The
symmetry approach to classification of integrable equations, in
What is integrability? (V.E. Zakharov, Ed.) pp. 115-184, Springer
series in Nonlinear Dynamics, 1991.

\bibitem{Ibragimov2} N.H. Ibragimov, Transformation groups applied to
mathematical physics (Dordrecht:Reidel), 1985.

\bibitem{KZ} E.A. Zabolotskaya and R.V. Khokhlov, Quasi-plane waves in the nonlinear acoustics of confined beams, Sov. Phys. Acoust. {\bf 15} (1969) 35--40.




\bibitem{Zakharov}  V.E. Zakharov,  Dispersionless limit of
integrable systems in $2+1$ dimensions, in Singular Limits of
Dispersive Waves, Ed. N.M. Ercolani et al., Plenum Press, NY
(1994) 165--174.


\bibitem{Shulman} V.E. Zakharov and E.I. Schulman,  Integrability of nonlinear systems and perturbation theory, in:  What is integrability?, 185--250, Springer Ser. Nonlinear Dynam., Springer, Berlin, 1991.

\bibitem{Zenchuk} A.I.  Zenchuk, The spectral problem and particular solutions to the
(2 + 1)-dimensional integrable generalization of the Camassa-Holm
equation, Physica D {\bf 152-153} (2001) 178--188.

\end{thebibliography}
\end{document}